\documentclass[useAMS,usenatbib]{mn2e}

\usepackage{graphicx}

\def\h0 {$h_0$=70 km s$^{-1}$ Mpc$^{-1}$}

\newcommand{\be}{\begin{equation}}
\newcommand{\ee}{\end{equation}}
\newcommand{\ce}{$\lambda_{\rm Edd}$}
\newcommand{\kms}{\ifmmode {\rm km\ s}^{-1} \else km s$^{-1}$\ \fi}
\newcommand{\tes}{\ifmmode \tau_{\rm es} \else $\tau_{\rm es}$\ \fi}
\newcommand{\tk}{\ifmmode \tau_{\rm K} \else $\tau_{\rm K}$\ \fi}
\newcommand{\vfwhm}{\ifmmode V {\mbox{\tiny FWHM}} \else
            $V {\mbox{\tiny FWHM}}$\fi}
\newcommand{\msun}{\ifmmode M_{\odot} \else $M_{\odot}$\ \fi}
\newcommand{\afe}{\ifmmode {\mathcal A {\rm Fe}} \else${\mathcal A {\rm Fe}}$\ \fi}

\newcommand{\lb}{\ifmmode L_{\rm bol} \else $L {\rm bol}$\ \fi}
\newcommand{\ledd}{\ifmmode L_{\rm Edd} \else $L {\rm Edd}$\ \fi}
\newcommand{\lx}{\ifmmode L_{\rm 2-10\ keV} \else  $L {\rm 2-10\ keV}$\ \fi}
\newcommand{\hb}{\ifmmode H\beta \else H$\beta$\ \fi}
\newcommand{\mbh}{\ifmmode M {\rm BH}  \else $M {\rm BH}$\ \fi}
\newcommand{\lv}{\ifmmode \lambda L {\lambda}(5100\AA) \else $\lambda L {\lambda}(5100\AA)$\ \fi}

\def\ka{K$\alpha$~}

\newcommand{\fei}{Fe {\sc i} \ }

\newcommand{\fexxv}{Fe {\sc xxv}\ }
\newcommand{\fexxvi}{Fe {\sc xxvi}\ }

\def\ariel5{{\it Ariel 5}\ }

\def\xmm{{\it XMM-Newton}~}
\def\chandra{{\it Chandra}}

\def\heao1{{\it HEAO~1}\ }

\def\rxte{{\it RXTE}\ }

\def\rxte{{\it RXTE}\ }
\def\suzaku{{\it Suzaku }\ }

\title[Fe K$\alpha$ line from accretion disks]{Ionization structure and Fe K$\alpha$ energy for irradiated accretion disks}
\author[X. L. Zhou, Y. H. Zhao \& R. Soria]{X. L. Zhou$^{1,2}$\thanks{E-mail: zhouxl@nao.cas.cn(XLZ)},
Y. H. Zhao$^{1,2}$ and R. Soria$^{3}$ \\
$^{1}$Key Laboratory of Optical Astronomy, National Astronomical
Observatories, Chinese Academy of Sciences, Beijing 100012, China \\
$^{2}$National Astronomical Observatories, Chinese Academy of
Sciences, Beijing, 100012, China \\
$^{3}$Mullard Space Science Laboratory, University College London,
Holmbury St Mary, Surrey RH5 6NT, UK}
\begin{document}

\date{Accepted 2011 . Received 2010; in original form 2010}

\pagerange{\pageref{firstpage}--\pageref{lastpage}} \pubyear{2009}

\maketitle

\label{firstpage}

\begin{abstract}

We study the radial ionization structure at the surface
of an X-ray illuminated accretion disk. We plot the expected
iron K$\alpha$ line energy as a function of the Eddington ratio
and of the distance of the emitting matter from the central source,
for a non-rotating and a maximally-rotating black hole.
We compare the predicted disk line energies with those
measured in an archival sample of active galactic nuclei observed
with {\it Chandra}, {\it XMM-Newton} and {\it Suzaku},
and discuss whether the line energies are consistent
with the radial distances inferred from reverberation studies.
We also suggest using rapidly-variable iron K$\alpha$ lines
to estimate the viscosity parameter of an accretion disk.
There is a forbidden region in the line energy versus Eddington
ratio plane, at low Eddington ratios, where an accretion disk
cannot produce highly-ionized iron K$\alpha$ lines.
If such emission is observed in low-Eddington-ratio
sources, it is either coming from a highly-ionized outflow, or
is a blue-shifted component from fast-moving neutral matter.
\end{abstract}


\section{Introduction}
\label{introduction} Recent high-resolution spectral studies by
\chandra, \xmm~and \suzaku have revealed that emission lines at
6.7 or 6.9 keV (probably emitted by highly ionized iron) are
common features in the X-ray spectra of active galactic nuclei
(AGNs, e.g.,
\citealt{Pounds2003,Reeves2004,Yaqoob2004,Bianchi2009,Shu2010,Patrick2010}).
The \ka emission lines produced by ionized iron in optically thin,
photoionized material in AGNs have been studied by
\citet[]{Bianchi2002}. Their results may require an iron
overabundance by a factor of a few to account for the observed line
strength.

The energy and ionization balance at the surface of an accretion disk
can change due to the X-ray illumination, and this affects
the line emission. Thus, it is important to determine the
radial ionization structure of X-ray photoionized accretion disks,
in order to study the reflected spectra and the associated
iron \ka emission. The energy of the iron \ka line generally increases
as iron becomes more stripped, from 6.4 keV for \fei to 6.97 keV for
\fexxvi (\citealt{Kaspi2002,Paerels2003}). This provides
a diagnostic of the accretion disk structure: we can compare
the calculated photoionization structure of the disk
with the observed energy of the iron \ka line, assuming
that the line was emitted from the disk.

In this letter, we calculate the radial ionization structure of an
X-ray illuminated AGN accretion disk as a function of Eddington ratio,
$\lambda_{\rm Edd} \equiv \lb/\ledd$, where $\lb$ is the bolometric
luminosity and $\ledd$ is the Eddington luminosity. We
then compare the predicted range of ionization parameters and
iron \ka line energies with a sample of recent line observations.
We plot the predicted and observed line energies $E_{\rm \alpha}$ as
a function of \ce. This gives us clues on the origin
of the iron \ka lines.
Throughout this Letter, we
assume standard cosmological parameters $h_0$ = 70 km s$^{-1}$ Mpc$^{-1}$,
$\Omega_{\rm m} = 0.27$, and $\Omega_{\rm \Lambda} = 0.73$.

\section{Methods and Results}
\label{sect:data} X-ray reflection from the surface of cold
matter around compact accreting objects has been studied by many
authors
(\citealt{Basko1974,Guilbert1988,Lightman1988,White1988,George1991,Matt1991,Done1992,
Ross1993,Matt1993,Czerny1994,Krolik1994,Magdziarz1995,Ross1996,Reynolds1997,Blackman1999,
Nayakshin2000,Young2000,Nayakshin2001,Ballantyne2002,Miniutti2004};
etc). Following those studies, we assume an idealized lamppost geometry
for the X-ray illumination of the reprocessing matter:
all the X-rays are emitted isotropically by a point source
located on the symmetry axis above a standard accretion disc
\citep[]{Shakura1973}, at a certain height $h$.
Ignoring the effects of light bending in the vicinity of the black hole (BH),
the ionization parameter is
\begin{equation}
\label{xi} \xi=8.97\times\left(\frac{\lambda_{\rm Edd}^3\eta_{\rm
x}\alpha}{\eta^2}\right)f^2(r)r^{-3/2}\,\frac{h}{\left(r^2+h^2\right)^{3/2}}
\end{equation}
\citep{Matt1993,Ross1993}, where
$r$ is the radial coordinate on the disk,
$f(r)=1-\sqrt{6/r}$,
$\eta$ is the total radiative efficiency,
$\eta_{\rm x}$ is the X-ray fraction ($L_{\rm x}\equiv \eta_{\rm x}$ \ce $\ledd$), and
$\alpha$ is the viscosity parameter.
In Eq.(1), $h$ and $r$ are in units of $r_{\rm g}\equiv GM_{\rm BH}/c^2$.
The efficiency $\eta=0.06$ for a Schwarzschild BH and
$\eta=0.31$ for a maximally spinning astrophysical BH \citep[]{Thorne1974}.
Here we assume a standard value of viscosity parameter of $\alpha\sim0.1$,
although a large range of values has been used in previous studies,
from $\alpha\sim0.01$ \citep[]{Miller2000,Starling2004}
to $\alpha\sim0.3$ \citep[]{Esin1997}.

We took the values of the iron K$\alpha$ line energy emitted by
an X-ray illuminated disk as a function of $\xi$ from
the calculations of \citet[][Fig 1b]{Matt1993}. Note
that the relation between $E_{\rm \alpha}$ and $\xi$
is almost independent of the incident angle of the illuminating flux;
here we assume that $E_{\rm \alpha}$ is only a function
of the iron ionization state. Using Eq.(1), we calculated
$E_{\rm \alpha}$ as a function of \ce~ for different values of $h$ and $r$
(Figure 1). Plotted as black lines are the results for a
Schwarzschild BH; the blue lines are the results for a maximally-spinning
Kerr BH. We repeated the calculation for six values of $r$
between $5r_{\rm g}$ and $200r_{\rm g}$ in the Schwarzschild case,
and the same six values of $r$ in the Kerr case. The radial values
are plotted as labels in Figure 1, next to the Schwarzschild curves.
For each value of $r$, we performed our calculation for three values
of $h=5r_{\rm g}$, $8r_{\rm g}$, and $15r_{\rm g}$. The curves
corresponding to those three values of $h$ are plotted
as solid, dashed and dot-dashed lines, respectively.

\begin{figure}
\includegraphics[width=58mm,angle=-90]{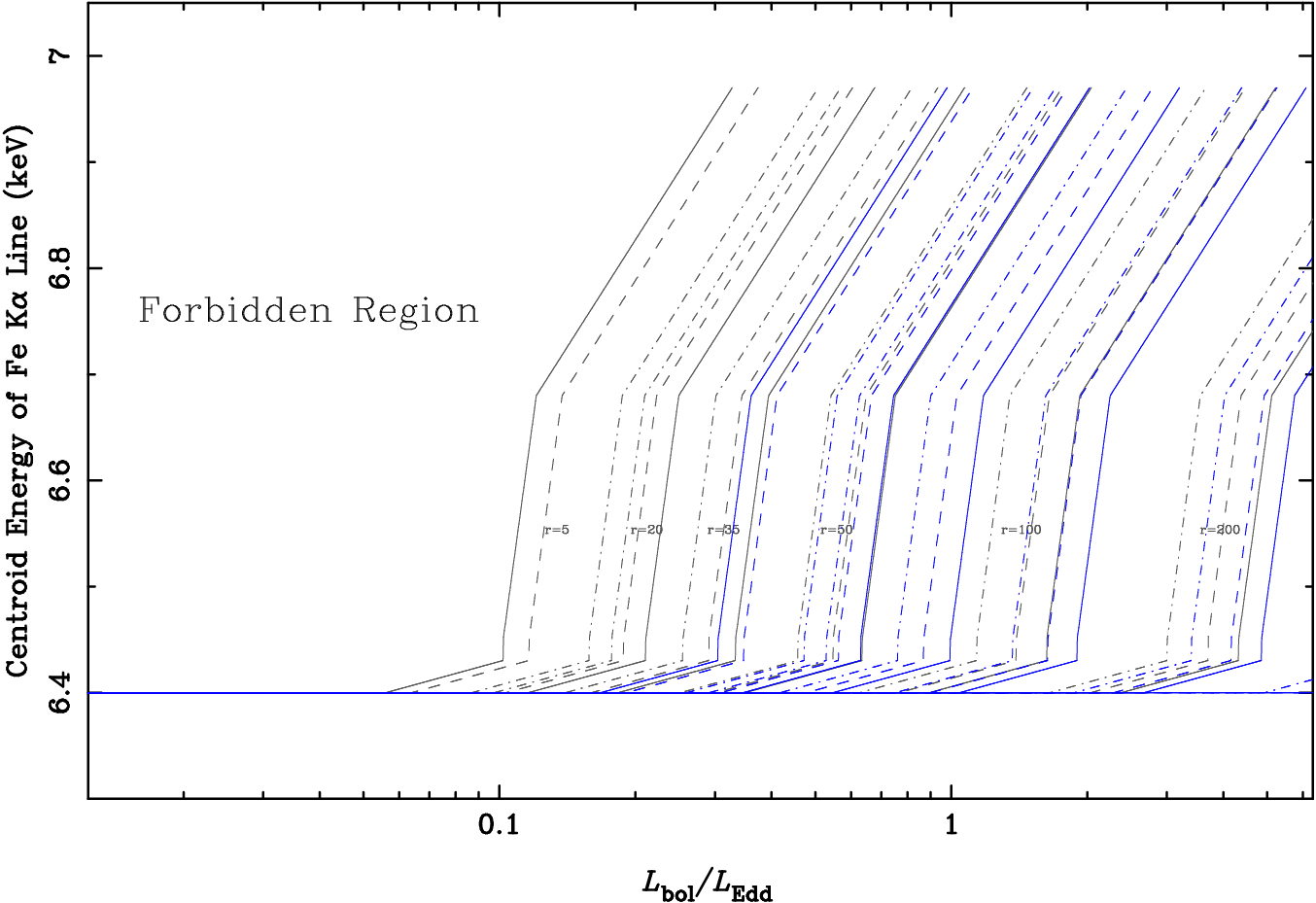}
\caption{Predicted line energies in the $(E_{\rm \alpha},\lambda_{\rm Edd})$ plane.
The black lines are the results for a Schwarzschild BH; the blue
lines are for a maximally spinning astrophysical BH. In each case, we repeated
the calculation for six values of $r$ (indicated as labels
next to the Schwarzschild curves), and three values of
$h=5r_{\rm g}$, $8r_{\rm g}$, and $15r_{\rm g}$.
For each $r$ value, the three corresponding $h$ curves are plotted
as solid, dashed and dot-dashed lines, respectively. }
\label{fig1}
\end{figure}

\begin{figure}
\includegraphics[width=58mm,angle=-90]{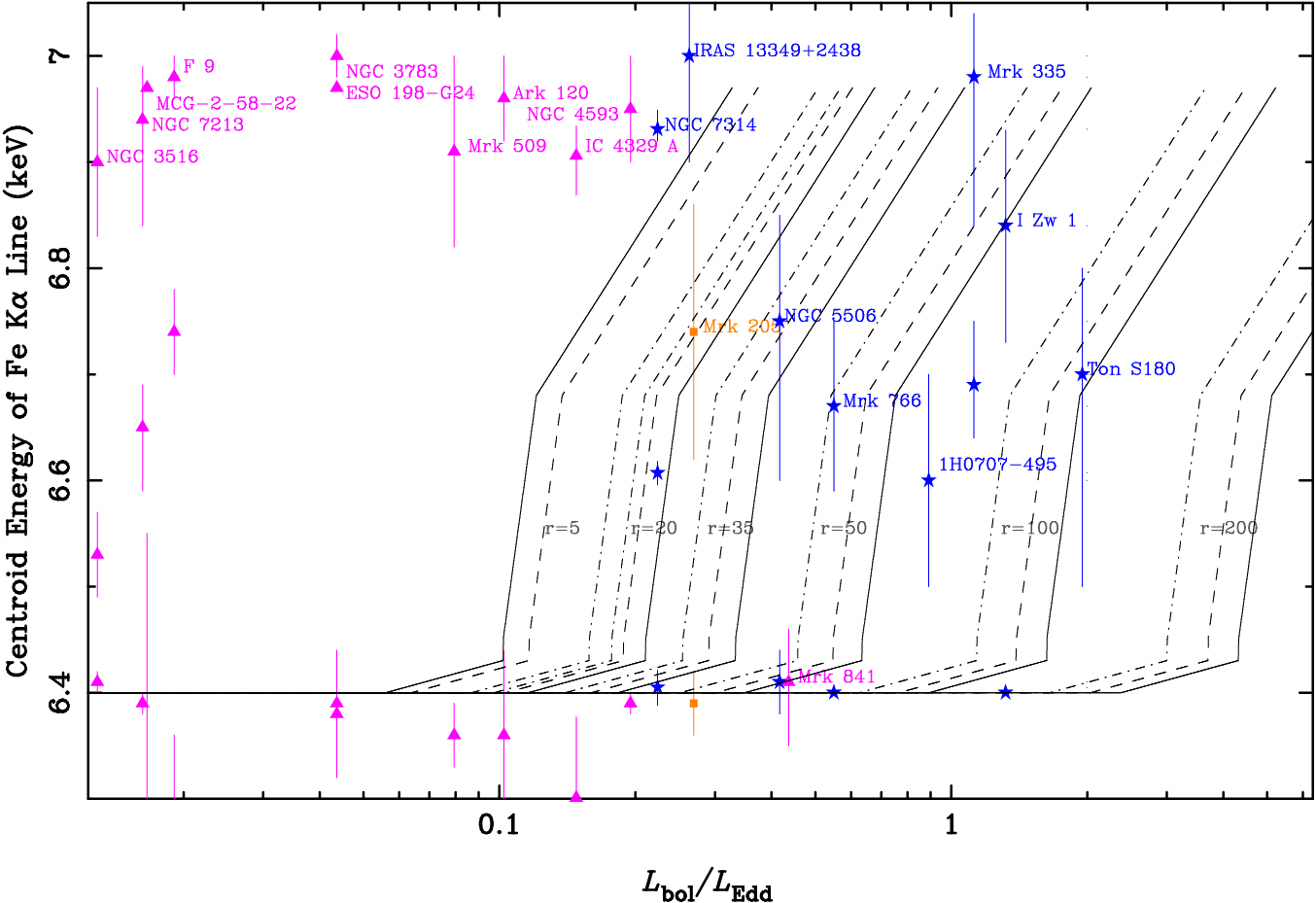}
 \caption{Comparison of the observed Fe K$\alpha$ line energies
with the predicted values. For simplicity, here we show only
the $(r,h)$ families of curves calculated for a Schwarzschild BH.
The radial distances and the meaning of the solid,
dashed and dot-dashed lines are the same as in Figure 1.
The $E_{\rm \alpha}$ values observed in narrow-line Seyfert 1 galaxies
(NLS1) are plotted as blue stars; those from broad-line Seyfert 1 galaxies
(BLS1) are magenta triangles; those from quasars are orange squares.
Multiple line energies are plotted for most galaxies. See Table 1
for details of the observed values.} \label{fig2}
\end{figure}

\section{Comparison with observations}
We compared the predicted and measured values of the line energy
in the $(E_{\rm \alpha},\lambda_{\rm Edd})$ plane (Figure 2; only
the calculations for a Schwarzschild BH are plotted, for simplicity).
The Eddington ratios \ce~ of the observed sample are taken from
\cite{Zhou2010} and \cite{Zhou2005}, who studied a large sample of X-ray luminous AGNs.
Using our theoretical curves, we can directly
estimate the effective distance of the line-emitting region
from the illuminating X-ray source, in the various sources.
For instance, observational values falling on the dashed line
corresponding to $(r=35r_{\rm g}, h=8r_{\rm g})$ indicate
a distance $R=(r^2+h^2)^{1/2} \approx 36~ r_{\rm g}$.
This distance determines a characteristic time lag between
primary X-ray source and line emission. From the observed
values of time lag and intrinsic line width, one can
estimate the BH mass and spin, with the iron line reverberation
method \citep[]{Reynolds1999,Liu2010}.
In the rest of this section, we compare the line emission
distances inferred from the $(E_{\rm \alpha},\lambda_{\rm Edd})$
plane with those estimated in the published literature, for a sample
of Seyfert galaxies.

{\sffamily NGC\,7314:} simultaneous observations of the NLS1 NGC\,7314
with \chandra ~and \rxte have revealed variability in the spectral features
on a timescale $< 12.5$ ks \citep[]{Yaqoob2003}, corresponding
to a light-crossing distance of $\approx 500 r_{\rm g}$ and a
Keplerian radius of $\approx 18.5 r_{\rm g}$ for a BH mass of $10^{6.70}
\msun$ \citep[]{Zhou2005}. This is in excellent agreement with the
{\fexxv} emission region ($\approx 20r_{\rm g}$) estimated
for this object from our line energy curves (Figure 2).

{\sffamily Mrk\,766:} the X-ray spectra of the NLS1 Mrk 766 show
a broad emission line at $\approx 6.7$ keV \citep[]{Pounds2003}.
From our plots in Figure 2, we infer that the \fexxv emission region
is located at $35r_{\rm g} \la r \la 50 r_{\rm g}$ from the central BH. In addition,
\cite{Turner2004} found a transient narrow line at $\approx 5.6$ keV
and interpreted it as evidence for blob ejection of neutral
or low-ionized material. If so, the rest-frame line energy
is $\approx 6.4$ keV, and the line emission region may be
located at $50 r_{\rm g} \la r \la 100 r_{\rm g}$, corresponding to a distance
of $\sim 10^{14}$ cm for a BH mass of $10^{6.6}$ \msun \citep[]{Zhou2005},
in broad agreement with the results of \cite{Turner2004}.

{\sffamily 1H0707$-$495:} the spectrum of this NLS1 galaxy shows a sharp
drop at energies $\approx 7$--$7.5$ keV\citep[]{Boller2002, Gallo2004b}.
A partial covering model was introduced to reduce the need
for an extreme iron overabundance \citep[]{Tanaka2004}.
However, an alternative possibility is that the X-ray spectrum
is dominated by ionized reflection rather than absorption.
In this scenario, it was argued \citep{Fabian2004} that the
two \xmm observations support the light bending model
of \citet[]{Miniutti2004}. \cite{Fabian2009} also measured
a time lag of about 30s  between the soft energy band ($0.3$--$1$ keV)
and the medium energy band ($1$--$4$ keV). Combining this short
time lag with the measured width of the broad iron L emission,
they suggested that the X-ray reverberation comes from matter
very close to the event horizon of a rapidly spinning BH.
However, considering a larger energy range ($0.3$--$7.5$ keV),
\cite{Miller2010} argued instead that the observed time delays
extend up to about 1800s in the hard band; this is consistent
with reverberation caused by scattering of X-rays passing
through much more distant absorbing material.
For our adopted mass of $10^{6.37}$ \msun \citep{Zhou2005},
$\lambda_{\rm Edd} \sim 1$.
This suggests that 1H0707$-$495 may have a highly-ionized
disk within $\approx 50 r_{\rm g}$ for a Schwarzschild BH,
or within $\approx 20 r_{\rm g}$ for a rapidly spinning BH.
In our scenario, those characteristic distances may be
the origin of the observed iron \ka lines at rest-frame energies
$\approx 6.5$--$6.7$ keV \citep[]{Fabian2009, Zoghbi2010}.

{\sffamily NGC\,3516:} simultaneous \chandra~ and \xmm~
observations showed \citep{Turner2002} two pairs of weak
emission features, including a component at $\approx 6.9$ keV,
symmetrically located around a strong, narrow 6.4 keV line.
This structure was interpreted \citep{Turner2002}
as evidence for relativistic broadening of disk lines
from three different radii. In our model, we find that
for $\lambda_{\rm Edd} \sim 0.01$ (Figure 2) even the innermost
region of the accretion disk must remain neutral.
Thus, if the observed $6.5$ keV and $6.9$ keV lines are emitted
from the disk, they must be blue-shifted
peaks of a $6.4$ keV line, rather than being emitted
by ionized iron. This supports the interpretation of \cite{Turner2002}.

{\sffamily Mrk\,841:} we infer that the observed $6.4$ keV line is emitted
from a region $\approx 50 r_{\rm g}$ from the central Schwarzschild BH
(or closer, for a Kerr BH). This corresponds to a time lag of 20 ks
for a BH mass of $10^{7.90}$ \msun \citep{Zhou2005}, in good agreement
with the variability timescale $<36$ ks found in two \xmm observations
\citep{Petrucci2002,Longinotti2004}.

\begin{table*}
\centering \caption{Observed rest-frame central energies of
the iron \ka lines in our sample of galaxies. Col.(1): common name of
the object; Col.(2): redshift,  from the NASA/IPAC
Extragalactic Database (NED); Col.(3): source of the X-ray data;
Col.(4--6): centroid energies ($^f$ fixed); Col.(7): Eddington ratio, from
Zhou \& Zhang (2010) and Zhou \& Wang (2005); Col.(8):
references for the X-ray spectra: (1) Patrick et
 al. (2010); (2) Gallo et al. (2004a); (3) Takahashi et al. (2010);
(4) Schmoll et al. (2009); (5) Bianchi et al. (2004); (6) Bianchi
 et al. (2005); (7) Fabian et al. (2009); (8) Turner et al. (2002);
(9)  Reeves et al. (2004); (10) Pounds et al. (2003); (11)  Reeves et
 al. (2001); (12) Reynolds et al. (2004); (13) Longinotti et al. (2003);
(14) McKernan \& Yaqoob (2004); (15) Matt et al. (2001); (16) Pounds et
 al. (2001); (17) Bianchi et al. (2003); (18) Yaqoob et al. (2003); (19)
 Petrucci et al. (2002).
 \label{tab:sample}}
\begin{center}
\begin{tabular}{lccccccr}
\hline
Source  & Redshift & Mission & $E_{\alpha1}$ & $E_{\alpha2}$ & $E_{\alpha3}$ & log(\ce) & Ref. \\
      &          &      &  (keV)        &  (keV)        &  (keV)
             &      & \\
(1) & (2) & (3) & (4) & (5) & (6) &  (7) & (8) \\
\hline
Mrk 335     & 0.0258 & \suzaku & $6.27^{+0.13}_{-0.17}$   & $6.69^{+0.06}_{-0.05}$ & $6.98^{+0.06}
_{-0.14}$   &   0.05  & 1 \\
I Zw 1      & 0.0589 & {\it XMM-Newton} & 6.4$^f$  & $6.84^{+0.09}_{-0.11}$ & ...
           &0.12 & 2 \\
Ton S180    & 0.0620 & \suzaku &...  & $6.7^{+0.1}_{-0.2}$ & ... & 0.29 & 3  \\
Fairall 9   & 0.0470 & \suzaku & $6.16\pm{0.2}$  &
         $6.74^{+0.04}_{-0.04}$ &  $6.98\pm0.02$   & $-1.72$  &1,4 \\
ESO 198-G24 &0.0455  & \chandra, {\it XMM-Newton}  &  $6.38\pm0.06$   & ...      &
             6.97$^f$  & $-1.36$ & 5,6 \\
Ark 120     & 0.0327 &  \suzaku  & $6.36^{+0.08}_{-0.09}$ & ... &
             $6.96\pm0.04$ & $-0.99$ &  1       \\
1H 0707-495 & 0.0406 & {\it XMM-Newton}    & ...  &  $6.5-6.7$  & ...  & $-0.04$ & 7   \\
NGC 3516    & 0.0089 & \chandra, {\it XMM-Newton}   & $6.41\pm0.01$ & $6.53\pm0.04$
         & $6.84-6.97$  & $-1.89$ & 8 \\
NGC 3783    & 0.0097 & {\it XMM-Newton}    & $6.39\pm{0.01}$   &... & $7.00\pm0.02$
            & $-1.36$ &9 \\
Mrk 766     & 0.0129 & {\it XMM-Newton}    & 6.40$^f$  & $6.67\pm0.08$  & ... &
             $-0.26$ & 10 \\
Mrk 205     &  0.0708& {\it XMM-Newton}    & $6.39\pm0.03$ & $6.74\pm0.12$ & ... &
             $-0.57$& 11  \\
NGC 4593    & 0.0090 & {\it XMM-Newton}    & $6.39\pm0.01$ &... & $6.95\pm0.05$ &
             $-0.71$  & 12 \\
IRAS 13349+2438 &  0.1076 & {\it XMM-Newton} & $6.0^{+0.3}_{-0.2}$ & ... &
             $7.0\pm0.1$ & $-0.58$ &  13 \\
IC 4329A    & 0.0161 &\chandra  & $6.301^{+0.076}_{-0.073} $  & ... &
         $6.906^{+0.028}_{-0.037}$ &  $-0.83$  & 14   \\
NGC 5506    & 0.0062 & {\it XMM-Newton}   & $6.41\pm0.03$  & $6.75^{+0.10}_{-0.15}$
         & ... & $-0.38$ & 15    \\
Mrk 509     & 0.0344 & {\it XMM-Newton}    &   $6.36\pm0.03$ & ... & $6.91\pm0.09$
             & $-1.10$     & 16    \\
NGC 7213    & 0.0058 & {\it XMM-Newton}    &     $6.39\pm0.01$  &
         $6.65^{+0.04}_{-0.06}$ & $6.94^{+0.05}_{-0.10}$&$-1.79$ &
                 17  \\
NGC 7314    &  0.0048 &\chandra & $6.405^{+0.016}_{-0.017}$ & $6.607^{+0.011}
_{-0.017}$ & $6.931^{+0.018}_{-0.011}$ & $-0.65$ & 18    \\
MCG-02-58-22&0.0469    &  {\it XMM-Newton} & $6.29^{+0.26}_{-0.06}$ & ... &
             6.97$^f$ & $-1.78$ & 5,6   \\
Mrk 841     & 0.0364 & {\it XMM-Newton}   &   $6.41^{+0.05}_{-0.06}$ & ... & ...
             &   $-0.36$   &   19  \\
\hline
\end{tabular}
\end{center}
\end{table*}

\section{discussion and conclusions}
\label{sect:concls}

Iron at the surface of an accretion disk is significantly
ionized when the Eddington ratio \ce~ is larger than a critical value.
Assuming a standard viscosity parameter $\alpha \sim0.1$,
the critical value above which iron in the innermost part of the disk
becomes ionized is $\lambda_{\rm Edd} \sim 0.1$ for a Schwarzschild BH,
and $\lambda_{\rm Edd} \sim 0.3$  for a maximally-rotating astrophysical BH.
We studied the radial ionization structure of an X-ray illuminated accretion disk,
and calculated the energy (increasing with the ionization parameter)
of the iron K$\alpha$ lines emitted from the disk.
We plotted those energies as family of curves in the
$(E_{\rm \alpha},\lambda_{\rm Edd})$ plane,
parameterized in terms of radial distance of the emitters
and height of the illuminating X-ray source above the disk plane
(lamppost model), for a non-rotating and a maximally-rotating BH.
We compared our model with the observed K$\alpha$ line energies
from a sample of AGNs.

A substantial fraction of AGNs show highly-ionized
iron \ka emission. The origin of the ionized emission is still debated.
Our results suggest that it may come from two different sources:
the accretion disk (for $\lambda_{\rm Edd} \ga 0.1$) or the photoionized material
in the outflow (for $\lambda_{\rm Edd} \la 0.1$).
Our model presented here is based on simple assumptions,
such as constant density without vertical stratification \citep{Matt1993},
but our main goal is to illustrate an important physical effect,
which is unlikely to depend substantially on the details
of the disk structure.

The critical \ce~ depends on $\alpha$, which parameterizes our ignorance
of detailed accretion physics \citep{Ji2006,Miller2006}.
Despite forty years of observational, experimental and theoretical studies since
\citet{Shakura1973}, we are still unable to determine
the disk viscosity accurately. The theoretical dependence
of the observed iron K$\alpha$ line energy on $\alpha$ suggests
that we can reverse the argument: if we have independent
measurements of a BH mass, spin and luminosity,
we can estimate $\alpha$ using the ionization curves
in the $(E_{\rm \alpha},\lambda_{\rm Edd})$ plane,
by combining the information on centroid energy
and rapid variability timescale. It is plausible
that $\alpha\sim0.1$ is in agreement of iron line observations of
a few AGNs.

Iron near the disk surface cannot be ionized at low accretion rates
and low Eddington ratios. There is a forbidden region in the
$(E_{\rm \alpha},\lambda_{\rm Edd})$ plane, below which ionized
K$\alpha$ line emission cannot come from an irradiated disk.
Observationally, several low-luminosity AGNs in that region show
K$\alpha$ emission features at $6.5$--$6.9$ keV. We argued that such
features are either coming from a highly-ionized outflow, or are
blue-shifted components from fast-moving neutral matter.
Alternatively, the intermediate energy line ($6.5$--$6.7$ keV)
seen in NGC\,3516, NGC\,7213 and Fairall 9 may come
from an evaporating/condensating region, as predicted by
the disk transition model at $\lambda_{\rm Edd} \sim 0.01$--$0.02$
\citep[]{Czerny2000, Liu2009, Qiao2009}.

Finally, we emphasize that the current observations do not yet
allow us to put robust constraints on the origin of
the highly-ionized iron \ka lines. The line parameters derived
from X-ray spectral fitting are strongly model-dependent,
and have large uncertainties. Future observations
with the next generation of X-ray space telescopes (such
as the proposed {\it GRAVITAS} mission) will
resolve the profile and constrain the origin of those lines,
and test X-ray reverberation mapping in a large sample of AGNs.

\section*{Acknowledgements}
We are very grateful to an anonymous referee for
helpful comments to improve the manuscript substantially.
We thank the discussion and suggestion from Prof. J. M., Wang. This
work is supported by the National Natural Science Foundation of
China under grant 11003022 and the Guoshoujing Telescope. The
Guoshoujing Telescope (formerly named the Large Sky Area
Multi-Object Fiber Spectroscopic Telescope; LAMOST) is funded by the
National Development and Reform Commission, operated and managed by
the Key Laboratory of Optical Astronomy, NAOC, CAS. This research
has made use of results obtained with \chandra, \xmm and {\it
Suzaku}, which are collaborative missions contributed by the USA
(NASA), the ESA member states and the space agencies of Japan
(JAXA).

\label{lastpage}

\begin{thebibliography}{}
\bibitem[\protect\citeauthoryear{Ballantyne et~al.}{2002}]{Ballantyne2002}
Ballantyne D. R., Fabian A. C., Ross R. R., 2002, MNRAS, 329, L67
\bibitem[\protect\citeauthoryear{Basko et~al.}{1974}]{Basko1974}
Basko M. M., Sunyaev R. A., Titarchuk L. G., 1974, A\&A, 31, 249
\bibitem[\protect\citeauthoryear{Bhayani \& Nandra}{2010}]{Bhayani10}
Bhayani S., Nandra K., 2010, MNRAS, 408, 1020
\bibitem[\protect\citeauthoryear{Bianchi \& Matt}{2002}]{Bianchi2002}
Bianchi S., Matt G., 2002, A\&A, 387, 76
\bibitem[\protect\citeauthoryear{Bianchi et~al.}{2003}]{Bianchi2003}
Bianchi S., Matt G., Balestra I., Perola G. C.,  2003, A\&A, 407, L21
\bibitem[\protect\citeauthoryear{Bianchi et~al.}{2004}]{Bianchi2004}
Bianchi S., Matt G., Balestra I., Guainazzi M., Perola G. C. 2004,
A\&A, 422, 65
\bibitem[\protect\citeauthoryear{Bianchi et~al.}{2005}]{Bianchi2005}
Bianchi S., Matt G., Nicastro F., Porquet D., Dubau J. 2005, MNRAS, 357,599
\bibitem[\protect\citeauthoryear{Bianchi et~al.}{2009}]{Bianchi2009}
Bianchi S., Guainazzi M., Matt G., Fonseca B. N., Ponti G., 2009,
A\&A, 495, 421
\bibitem[\protect\citeauthoryear{Blackman}{1999}]{Blackman1999}
Blackman E. G., 1999, MNRAS, 306, L25
\bibitem[\protect\citeauthoryear{Boller et~al.}{2002}]{Boller2002}
Boller T. et~al., 2002, MNRAS, 329, L1
\bibitem[\protect\citeauthoryear{Czerny \& \.Zycki}{1994}]{Czerny1994}
Czerny B.,  \.Zycki P. T., 1994, ApJ, 431, L5
\bibitem[\protect\citeauthoryear{Done et~al.}{1992}]{Done1992}
Done C., Mulchaey J. C., Mushotzky R. F., Arnaud K. A., 1992, ApJ,
395, 275
\bibitem[\protect\citeauthoryear{Esin et~al.}{1997}]{Esin1997}
Esin A. A., McClintock J. E.,  Narayan R., 1997, ApJ, 489, 865
\bibitem[\protect\citeauthoryear{Fabian et~al.}{2004}]{Fabian2004}
Fabian A. C., Miniutti, G., Gallo L., Boller T., Tanaka Y., Vaughan
S., Ross R. R., 2004, MNRAS, 353, 1071
\bibitem[\protect\citeauthoryear{Fabian et~al.}{2009}]{Fabian2009}
Fabian A. C. et~al., 2009, Nature, 459, 540
\bibitem[\protect\citeauthoryear{Gallo et~al.}{2004a}]{Gallo2004a}
Gallo L. C., Tanaka Y., Boller T., Fabian A. C., Vaughan S., Brandt
W. N., 2004, A\&A, 417, 29
\bibitem[\protect\citeauthoryear{Gallo et~al.}{2004b}]{Gallo2004b}
Gallo L. C., Tanaka Y., Boller T., Fabian A. C., Vaughan S., Brandt
W. N., 2004, MNRAS, 353, 1064
\bibitem[\protect\citeauthoryear{George \& Fabian}{1991}]
{George1991} George I. M., Fabian A. C., 1991, MNRAS, 249, 352
\bibitem[\protect\citeauthoryear{Guilbert \&
Rees}{1988}]{Guilbert1988} Guilbert P. W., Rees M. J., 1988, MNRAS,
233, 475
\bibitem[\protect\citeauthoryear{Kaspi et~al.}{2002}]{Kaspi2002}
Kaspi S. et~al., 2002, ApJ, 574, 643
\bibitem[\protect\citeauthoryear{Krolik et~al.}{1994}]{Krolik1994}
Krolik J. H., Madau P., \.Zycki P. T. 1994, ApJ, 420, L57
\bibitem[\protect\citeauthoryear{Ji et~al.}{2006}]{Ji2006}
Ji H., Burin M., Schartman E., Goodman J., 2006, Nature, 444, 343
\bibitem[\protect\citeauthoryear{Lightman \& White}{1988}]{Lightman1988}
Lightman A. P., White T. R., 1988, ApJ, 335, 57
\bibitem[\protect\citeauthoryear{Liu \& Taam}{2009}]{Liu2009}
 Liu B. F., Taam R. E., 2009, ApJ, 707, 233
\bibitem[\protect\citeauthoryear{Liu et~al.}{2010}]{Liu2010}
Liu Y. et~al., 2010, ApJ, 710, 1228
\bibitem[\protect\citeauthoryear{Longinotti et~al.}{2003}]{Longinotti2003}
Longinotti A. L., Cappi M., Nandra K., Dadina M., Pellegrini S., 2003,
A\&A, 410, 471
\bibitem[\protect\citeauthoryear{Longinotti et~al.}{2004}]{Longinotti2004}
Longinotti A. L., Nandra K., Petrucci P. O., O'Neill P. M., 2004,
MNRAS, 355, 929
\bibitem[\protect\citeauthoryear{Magdziarz \& Zdziarski}{1995}]{Magdziarz1995}
Magdziarz P., Zdziarski A. A., 1995, MNRAS, 273, 837
\bibitem[\protect\citeauthoryear{Matt et~al.}{1991}]{Matt1991} Matt
G., Perola G. C., Piro L., 1991, A\&A, 247, 25
\bibitem[\protect\citeauthoryear{Matt et~al.}{1993}]{Matt1993} Matt
G., Fabian A. C., Ross R. R., 1993, MNRAS, 262, 179 (M93)
\bibitem[\protect\citeauthoryear{Matt et~al.}{2001}]{Matt2001}
Matt G., Guainazzi M., Perola G. C., Fiore F., Nicastro F., Cappi M.,
        Piro L. 2001, A\&A, 377, L31
\bibitem[\protect\citeauthoryear{McKernan \& Yaqoob}{2004}]{McKernan}
        McKernan B., Yaqoob T., 2004, ApJ, 608, 157
\bibitem[\protect\citeauthoryear{Miller et~al.}{2006}]{Miller2006}
Miller J. M., Raymond J., Fabian A. C., Steeghs D., Homan, J., Reynolds
C., van der Klis M., Wijnands R., 2006, Nature, 441, 953
\bibitem[\protect\citeauthoryear{Miller et~al.}{2010}]{Miller2010}
Miller L., Turner T. J., Reeves J. N., Braito V., 2010, MNRAS, 408,
1928
\bibitem[\protect\citeauthoryear{Miller \& Stone}{2000}]{Miller2000}
Miller K. A., Stone J. M., 2000, ApJ, 534, 398
\bibitem[\protect\citeauthoryear{Miniutti \& Fabian}{2004}]{Miniutti2004}
Miniutti G., Fabian A. C., 2004, MNRAS, 349, 1435
\bibitem[\protect\citeauthoryear{Nayakshin et~al.}{2000}]{Nayakshin2000}
Nayakshin S., Kazanas D., Kallman T. R., 2000, ApJ, 537, 833
\bibitem[\protect\citeauthoryear{Nayakshin \& Kallman}{2001}]{Nayakshin2001}
Nayakshin S., Kallman T. R., 2001, ApJ, 546, 406
\bibitem[\protect\citeauthoryear{Paerels \& Kahn}{2003}]{Paerels2003}
Paerels F. B. S., Kahn S. M., 2003, ARA\&A, 41, 291
\bibitem[\protect\citeauthoryear{Patrick et~al.}{2010}]{Patrick2010}
Patrick A. R., Reeves J. N., Porquet D., Markowitz A. G., Lobban A.
P., Terashima Y., 2010, MNRAS, ArXiv Astrophysics e-prints
\bibitem[\protect\citeauthoryear{Petrucci et~al.}{2002}]{Petrucci2002}
Petrucci P. O. et al., 2002, A\&A, 388, L5
\bibitem[\protect\citeauthoryear{Pounds et~al.}{2001}]{Pounds2001}
Pounds K. A., Reeves J. N., O'Brien P. T., Page K. L., Turner M.,
Nayakshin S., 2001, ApJ, 559, 181
\bibitem[\protect\citeauthoryear{Pounds et~al.}{2003}]{Pounds2003}
Pounds K. A., Reeves J. N., Page K. L., Wynn G. A., O'Brien P. T.,
2003, MNRAS, 342, 1147
\bibitem[\protect\citeauthoryear{Poutanen et~al.}{1996}]{Poutanen1996}
Poutanen J., Nagendra K. N., Svensson R., 1996, MNRAS, 283, 892
\'o\.z
\bibitem[\protect\citeauthoryear{Qiao \&  Liu}{2009}]{Qiao2009}
 Qiao E., Liu B. F. 2009, PASJ, 61, 403
\bibitem[\protect\citeauthoryear{Reeves et~al.}{2001}]{Reeves2001}
Reeves J. N., Turner M. J. L., Pounds K. A., O'Brien P. T., Boller Th.,
        Ferrando P., Kendziorra E., Vercellone S. 2001,  A\&A,
        365, L134
\bibitem[\protect\citeauthoryear{Reeves et~al.}{2004}]{Reeves2004}
Reeves J. N., Nandra K., George I. M., Pounds K. A., Turner T. J.,
        Yaqoob T., 2004, ApJ, 602, 648
\bibitem[\protect\citeauthoryear{Reynolds \& Begelman}{1997}]{Reynolds1997}
Reynolds C. S., Begelman M. C., 1997, ApJ, 488, 109
\bibitem[\protect\citeauthoryear{Reynolds et~al.}{1999}]{Reynolds1999}
Reynolds C. S., Young A. J., Begelman M. C., Fabian A. C., 1999,
ApJ, 514, 164
\bibitem[\protect\citeauthoryear{Reynolds et~al.}{2004}]{Reynolds2004}
Reynolds C. S., Brenneman L. W., Wilms J., Kaiser M. E. 2004, MNRAS,
        352, 205
\bibitem[\protect\citeauthoryear{R\'o\.za\'nska \& Czerny}{2000}]
{Czerny2000} R\'o\.za\'nska A., Czerny B., 2000, A\&A, 360, 1170
\bibitem[\protect\citeauthoryear{Ross \& Fabian}{1993}]{Ross1993}
 Ross R. R., Fabian A. C., 1993, MNRAS, 261, 74
\bibitem[\protect\citeauthoryear{Ross et~al.}{1996}]{Ross1996}
Ross R. R., Fabian A. C., Brandt W. N., 1996, MNRAS, 278, 1082
\bibitem[\protect\citeauthoryear{Shakura \& Sunyaev}{1973}]{Shakura1973}
Shakura N. I., Sunyaev R. A., 1973, A\&A, 24, 337
\bibitem[\protect\citeauthoryear{Schmoll et~al.}{1999}]{Schmoll1999}
Schmoll S. et al., 2009, ApJ, 703, 2171
\bibitem[\protect\citeauthoryear{Shu et~al.}{2010}]{Shu2010}
Shu X. W., Yaqoob T., Wang J. X., 2010, ApJS, 187, 581
\bibitem[\protect\citeauthoryear{Starling et al.}{2004}]{Starling2004}
Starling R. L. C., Siemiginowska A., Uttley P., Soria R., 2004,
   MNRAS, 347, 67
\bibitem[\protect\citeauthoryear{Takahashi et~al.}{2010}]{Takahashi2010}
Takahashi H., Hayashida K., Anabuki N., 2010, PASJ, 62, 1483
\bibitem[\protect\citeauthoryear{Tanaka et~al.}{2004}]{Tanaka2004}
Tanaka Y., Boller T., Gallo L., Keil R., Ueda Y., 2004, PASJ, 56, 9
\bibitem[\protect\citeauthoryear{Thorne }{1974}]{Thorne1974}
Thorne K. S., 1974, ApJ, 191, 507
\bibitem[\protect\citeauthoryear{Turner et~al.}{2002}]{Turner2002}
Turner T. J. et al., 2002, ApJ, 574, 123
\bibitem[\protect\citeauthoryear{Turner et~al.}{2004}]{Turner2004}
Turner T. J., Kraemer S. B., Reeves J. N., 2004, ApJ, 603, 62
\bibitem[\protect\citeauthoryear{White et~al.}{1988}]{White1988}
White T. R., Lightman A. P., Zdziarski A. A., 1988, ApJ, 331, 939
\bibitem[\protect\citeauthoryear{Yaqoob et~al.}{2003}]{Yaqoob2003}
Yaqoob T., George I. M., Kallman T. R., Padmanabhan U., Weaver K.
A., Turner T. J. 2003, ApJ, 596, 85
\bibitem[\protect\citeauthoryear{Yaqoob \& Padmanabhan}{2004}]{Yaqoob2004}
Yaqoob T., Padmanabhan U., 2004, ApJ, 604, 63
\bibitem[\protect\citeauthoryear{Young \& Reynolds}{2000}]{Young2000}
Young A. J., Reynolds C. S., 2000, ApJ, 529, 101
\bibitem[\protect\citeauthoryear{Zhou \& Wang}{2005}]{Zhou2005} Zhou X. L.,  Wang J. M., 2005, ApJ, 618, L83
\bibitem[\protect\citeauthoryear{Zhou \& Zhang}{2010}]{Zhou2010} Zhou X. L.,  Zhang S. N., 2010, ApJ, 713, L11
\bibitem[\protect\citeauthoryear{Zoghbi et~al.}{2010}]{Zoghbi2010}
 Zoghbi A., Fabian A. C., Uttley P., Miniutti G., Gallo L. C., Reynolds C. S., Miller J. M., Ponti
 G., 2010, MNRAS, 401, 2419
\end{thebibliography}
\end{document}